%% file: dutycycle.tex
\documentclass[iop]{emulateapj}

\newcommand{\chandra} {{\it Chandra}}
\newcommand{\nustar} {{\it NuSTAR}}

\newcommand{\cmsq} {cm$^{-2}$}
\newcommand{\nh} {$N_{\rm{H}}$}
\newcommand{\lx} {$L_{\rm{X}}$}

\newcommand{\chisq} {$\chi^2$}

\newcommand{\rchisq} {$\chi^2_r$}

\newcommand{\hii}{{\rm{H\,\sc{ii}}}}

\newcommand{\ergs}{\mbox{\thinspace erg\thinspace s$^{-1}$}}
\newcommand{\ergcms}{\mbox{\thinspace erg\thinspace cm$^{-2}$\thinspace s$^{-1}$}}

\newcommand{\msol} {$M_{\odot}$}

\shorttitle{Chandra and NuSTAR observations of the ULX pulsar in M82.}
\shortauthors{Brightman et al.}

\begin{document}

\title{Spectral and temporal properties of the ultra-luminous X-ray pulsar in M82 from 15 years of Chandra observations and analysis of the pulsed emission using NuSTAR}

\author{Murray Brightman$^{1}$, Fiona Harrison$^{1}$, Dominic J. Walton$^{2,1}$, Felix Fuerst$^{1}$, Ann Hornschemeier$^{3,4}$, Andreas Zezas$^{5,6,7}$, Matteo Bachetti$^{8}$, Brian Grefenstette$^{1}$, Andrew Ptak$^{3,4}$, Shriharsh Tendulkar$^{1}$, Mihoko Yukita$^{3,4}$}

\affil{$^{1}$Cahill Center for Astrophysics, California Institute of Technology, 1216 East California Boulevard, Pasadena, CA 91125, USA\\
$^{2}$Jet Propulsion Laboratory, California Institute of Technology, Pasadena, CA 91109, USA\\
$^{3}$NASA Goddard Space Flight Center, Code 662, Greenbelt, MD 20771, USA\\
$^{4}$The Johns Hopkins University, Homewood Campus, Baltimore, MD 21218, USA\\
$^{5}$Physics Department, University of Crete, Heraklion, Greece\\
$^{6}$Harvard-Smithsonian Center for Astrophysics, 60 Garden Street, Cambridge, MA 02138, USA\\
$^{7}$Foundation for Research and Technology-Hellas, 71110 Heraklion, Crete, Greece\\
$^{7}$INAF/Osservatorio Astronomico di Cagliari, via della Scienza 5, I-09047 Selargius (CA), Italy\\}

\begin{abstract}

The recent discovery by \cite{bachetti14} of a pulsar in M82 that can reach luminosities of up to $10^{40}$ \ergs, a factor of $\sim100$ the Eddington luminosity for a 1.4 \msol\ compact object, poses a challenge for accretion physics. In order to better understand the nature of this source and its duty cycle, and in the light of several physical models that have been subsequently published, we conduct a spectral and temporal analysis of the 0.5-8 keV X-ray emission from this source from 15 years of \chandra\ observations. We analyze 19 ACIS observations where the PSF of the pulsar is not contaminated by nearby sources. We fit the \chandra\ spectra of the pulsar with a power-law model and a disk black body model, subjected to interstellar absorption in M82. We carefully assess for the effect of pile-up in our observations, where 4 observations have a pile-up fraction $>10\%$, which we account for during spectral modeling with a convolution model.  When fitted with a power-law model, the average photon index when the source is at high luminosity (\lx$>10^{39}$ \ergs) is $\Gamma=1.33\pm0.15$. For the disk black body model, the average temperature is $T_{\rm in}=3.24\pm0.65$ keV, the spectral shape being consistent with other luminous X-ray pulsars. We also investigated the inclusion of a soft excess component and spectral break, finding that the spectra are also consistent with these features common to luminous X-ray pulsars. In addition, we present spectral analysis from \nustar\ over the 3$-$50 keV range where we have isolated the pulsed component. We find that the pulsed emission in this band is best fit by a power-law with a high-energy cut-off, where $\Gamma=0.6\pm0.3$ and $E_{\rm C}=14^{+5}_{-3}$ keV. While the pulsar has previously been identified as a transient, we find from our longer-baseline study that it has been remarkably active over the 15-year period, where for 9/19 (47\%) observations that we analyzed, the pulsar appears to be emitting at a luminosity in excess of $10^{39}$ \ergs, greater than 10 times its Eddington limit.

\end{abstract}


\section{Introduction}

The discovery of coherent pulsations with a period of 1.37 s in the X-ray emission of M82 by {\nustar} \citep[][hereafter B14]{bachetti14}, shown to be associated with an ultra-luminous X-ray source (ULX) that is known to reach luminosities of $10^{40}$ \ergs\ (with the assumption that the source radiates isotropically) is a challenge to accretion physics and has fuelled speculation as to the nature of this source \citep{lyutikov14, christodoulou14, fragos15, eksi15, kluzniak15, shao15, dallosso15, mushtukov15}. Since the pulsations are almost certainly produced by a rapidly spinning magnetized neutron star with a mass of $\sim$1.4 \msol, the observed peak X-ray luminosity is 100 times the system's Eddington limit. 

From B14, the neutron star orbits its companion with a 2.5-day period that is close to circular, having a projected semi-major axis of 22.225 light s (6.66$\times10^{6}$ km) and a companion star with a minimum mass of 5.2 \msol. A linear spin up is also observed from the pulsations during the \nustar\ observations, with a pulse derivative $\dot{P}\simeq-2\times10^{-10}$ s s$^{-1}$, that varies from observation to observation. The pulse profile is also close to sinusoidal (B14). Any theoretical model must be able to account for all of these properties.

High $B$-field ($B\gtrsim10^{12}$ G) accreting pulsars have been observed to emit in excess of their isotropic Eddington luminosities. Magnetic fields allow pulsars to exceed their Eddington luminosities by funneling the accreting material along the magnetic field lines onto the magnetic poles of the neutron star, with the X-ray emission radiating out the sides of the accretion column. Observational evidence that showed the pulsar SMC X-1 to be super-Eddington motivated calculations of radiative transfer in the presence of these magnetic fields by \cite{basko75}. The authors calculate the limiting luminosity of these systems, showing that this depends strongly on the geometry of the accretion channel \citep{basko76}. \cite{dallosso15} consider the observational properties of the pulsar in M82 and numerically solve the torque equation with respect to the scenario in which matter is funneled along the magnetic field lines, and argue in favor of a high magnetic field strength (B$\sim10^{13}$ G) to explain the variation in $\dot{P}$ and the pulsar's high luminosity. In this case the strength of the magnetic field would disrupt the accretion disk at much larger radii ($\lesssim(80-90) R_{\rm NS}$), causing the disk temperature to decrease. This would be difficult to observe, however, since the X-ray emission from the accreting pulsar is dominated by the base of the accretion column near the neutron star's surface. A magnetic field of B$\sim10^{13}$ G or stronger was also favored for the ULX pulsar by \cite{eksi15}, also based on calculations of torque equilibrium. Magnetic fields of these strengths have the power to reduce the Compton scattering cross section, thus increasing the critical accretion luminosity. \cite{dallosso15} calculate that for 30 keV photons, a $B\sim10^{13}$ G field would decrease the scattering cross section by a factor of $\sim50$.  

Conversely however, \cite{kluzniak15} argue for a low magnetic field strength (B$<10^{9}$ G). They base their conclusions on the ratio of spin-up rate to luminosity (10$^{-50}$ (erg s)$^{-1}$), which is an order of magnitude lower than typical X-ray pulsars. These authors argue that a disk truncated at large radii would not provide the required lever arm to power the observed spin-up. Another interpretation of the magnetic field by \cite{christodoulou14} implies that the magnetic field is in fact typical of pulsars at $\sim10^{12}$ G and that the observed luminosity can be accounted for by geometric beaming. However, the near-sinusoidal shape of the pulse profile suggests that strong beaming does not occur, and further, Eddington ratios of $\sim$100 are difficult to reconcile with this scenario. It is clear from these arguments that further observational evidence, such as the duty cycle of the source and X-ray spectral properties are needed to gain more insights into the nature of the source and its evolution. 

The ULX associated with the pulsar was first resolved by \chandra\ HRC in October 1999 \citep{matsumoto01} and designated CXOM82 J095551.1+694045. This source is now known to be the second most luminous X-ray source in M82 \citep{feng07, kong07} after the ULX M82 X-1 (CXOU J095550.2+694047) and thus is commonly referred to as M82 X-2. Both \cite{feng07} and \cite{kong07} use the \chandra\ data on M82 to study this source, noting its high X-ray luminosity and its month-timescale variability, identifying it as a transient. As for associations at other wavelengths, \cite{kong07} found that X-2 is coincident with the position of a star cluster seen in a near-infrared {\it HST} NICMOS F160W image, also associated with the radio source 42.21+59.2 from \cite{mcdonald02}, identified as an \hii\ region. Furthermore, \cite{gandhi11} presented high-resolution mid-infrared imaging of the center of M82 and tentatively assigned their source \#11 as a counterpart to X-2, based on its proximity to the position of the radio source. M82 X-1, which generally dominates the X-ray emission of the galaxy, is a candidate for an intermediate mass black hole, based on its extreme X-ray luminosity, which reach up to $10^{41}$ \ergs\ \cite[e.g.][]{kaaret06} and the detection of twin-peaked quasi-periodic oscillations at frequencies of 3.3 and 5.1 Hz \citep{pasham14}. M82 X-1 and X-2 are separated by only 5\arcsec\ and thus only resolvable by \chandra. 

Since the results presented in \cite{feng07} and \cite{kong07}, M82 has been observed by \chandra\ on 18 further occasions, including the \chandra\ data used in B14 to identify the source of the ultra-luminous pulsations. \cite{feng07} and \cite{kong07} identify X-2 as a transient, however, B14 found that the source retains its high luminosity 7 years later. If X-2 persists at luminosities of $\sim10^{40}$ \ergs, assuming a mass to energy conversion efficiency of unity, the neutron star will grow at a rate of $\sim2\times10^{-7}$ \msol\ yr$^{-1}$, meaning it will collapse into a black hole within $\sim10$ million years. 

The goal of this paper is to better understand the duty cycle of the source and its spectral characteristics, and discuss these characteristics with respect to theoretical models and other luminous pulsars. While only \chandra\ data can be used to spatially resolve the pulsar, \nustar, with its timing capabilities, allow it to temporally isolate the pulsed component due to the coherent pulsations emitted by the source. We carry out spectral analysis of the source in the 0.5$-$8 keV band using archival \chandra\ data and spectroscopy in the 3$-$50 keV band using \nustar. In Section \ref{sec_canalysis} we describe the \chandra\ data and analysis and in Section \ref{sec_nanalysis} we describe the \nustar\ data and analysis. In Section \ref{sec_results} we present our results and in Section \ref{sec_discuss} we discuss our findings with respect to other luminous X-ray pulsars and their implications for theoretical models. A distance of 3.3 Mpc to M82 is assumed throughout  \citep{foley14}.

\section{Chandra data and analysis}
\label{sec_canalysis}

In total, M82 has been observed 28 times by \chandra, first on 1999 September 20 with ACIS-I and most recently on 2015 January 20 with ACIS-S, covering more than 15 years of activity in the galaxy. The data contain a range of exposure times for each obsID, ranging from 2$-$120 ks, taken with all three instruments, ACIS-I, ACIS-S and HRC, all without the use of the gratings. Furthermore, the galaxy has been placed at a mixture of on-axis and off-axis angles, the off-axis angles being used to mitigate pileup from the bright X-ray sources by spreading out the counts over a wider area of the detector. This has often been combined with sub-array readout, typically 1/8th of one ACIS-S chip, also to mitigate pileup by reducing frame times from 3.2 s to 0.4 s. These observational data are summarized in Table \ref{table_obsdat}.

\begin{table*}
\centering
\caption{Chandra observational data}
\label{table_obsdat}
\begin{center}
\begin{tabular}{r c c c c l}
\hline
ObsID	& Date	& Instrument	& Exposure	& Pile-up fraction & Notes \\
	&	&	& (ks) \\
(1) & (2) & (3) & (4) & (5) & (6)  \\
\hline

\input{tab_obsdat.tex}

\hline
\end{tabular}
\tablecomments{Details of the 28 \chandra\ observations of M82 taken from 1999 to 2015, ordered by date. Column (1) gives the obsID (* indicates that this observation was not used in our investigation due to blended PSFs or HRC data), column (2) gives the date of the observation, column (3) gives the instrument used, column (4) gives the total exposure time in ks, column (5) gives an estimate of the pile-up fraction described in Section \ref{sec_canalysis} and column (6) gives details about if the observation was taken off-axis and/or with a sub-array of pixels in order to mitgate the effect of pile-up. We also note if the PSF of X-2 is blended with nearby sources.}
\end{center}
\end{table*}

\begin{figure*}
\begin{center}
\includegraphics[width=220mm,angle=90]{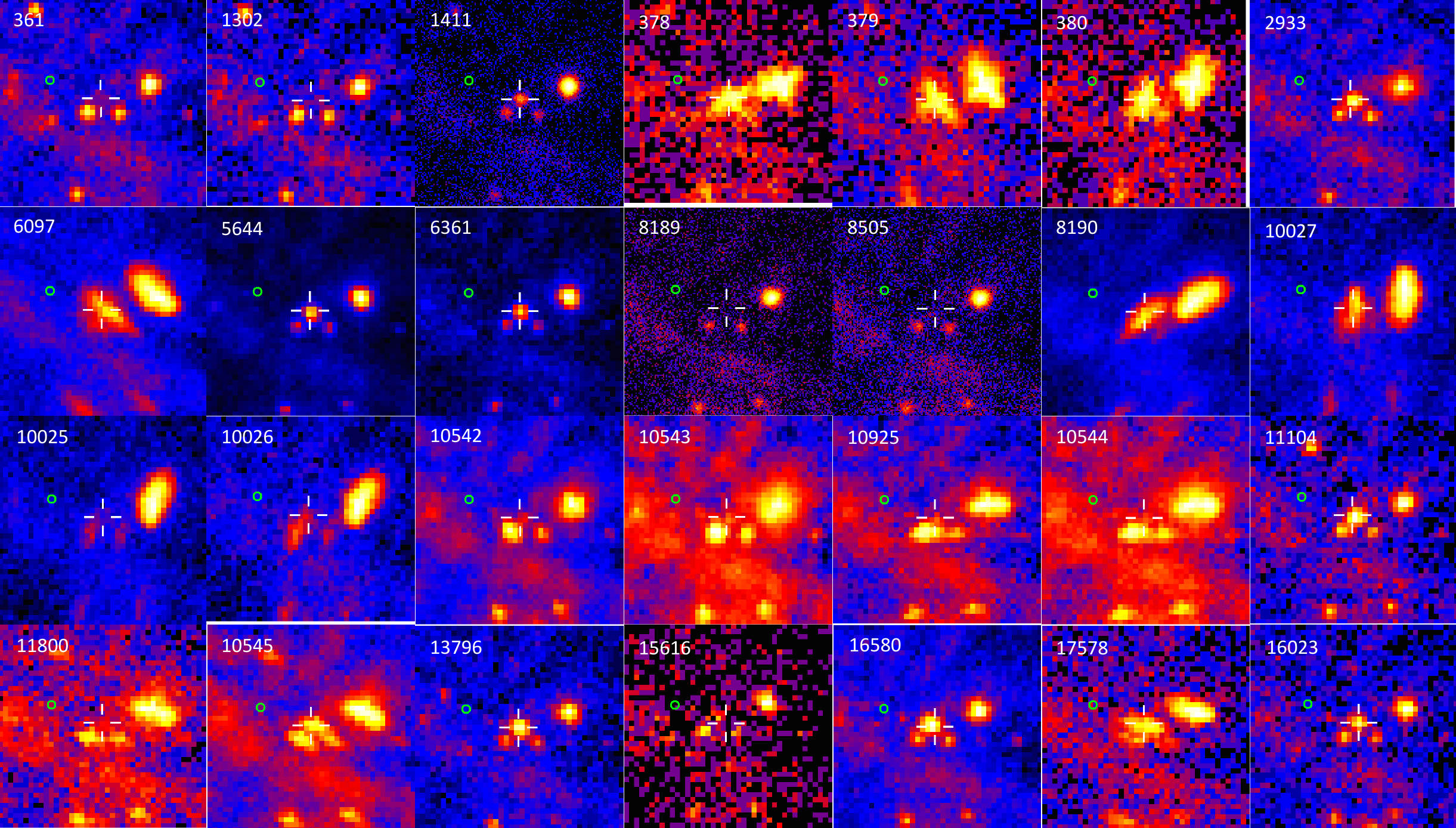}
\caption{20\arcsec$\times$20\arcsec\ 0.5$-$8 keV \chandra\ images of the central region of M82, consisting of 28 obsIDs taken over the 15 year-period 1999 to 2015. The obsIDs are written in the top left of each image. Images are centered on the position of the ultra-luminous pulsar, X-2, which is marked with white cross-hairs. The data were taken with a variety of instruments, exposure times, off-axis angles and detector sub-arrays, details of which are given in Table \ref{table_obsdat}. From top-left to bottom-right, the images are ordered by epoch, starting in September 1999 and ending in January 2015. The green circle to the north-east of X-2 marks the radio kinematic center of the galaxy from \cite{weliachew84}. North is to the top of the image and east is to the left.}
\label{fig_images}
\end{center}
\end{figure*}

We also present images of the central 20\arcsec$\times$20\arcsec\ region of M82 from each of the 28 obsIDs, centered on X-2 in Fig \ref{fig_images}. This figure also illustrates the range of \chandra\ data available on this galaxy. It is clear from this figure that not all the data can be used to study X-2, as in many cases the PSF of X-2 is blended with the PSF of two nearby sources to the south (i.e. obsIDs 378, 379, 380, 6097, 8190, and 10027).

We carry out analysis on level 2 event data that has undergone standard data processing by the \chandra\ X-ray Center (v8.4.4 for obsID 10025; v8.4.5 for 2933, 10026, 10542, 10543, 10544, 10545, 10925, 11104, 11800, 13796; v8.5.1.1 for 5644, 6361, 15616; v10.2.1 for 16580; v10.3.1 for 17578 and v10.3.3 for 16023). We reprocess obsIDs 361 and 1302 using the {\sc ciao} (v4.7, CALDB v4.6.5) script {\sc chandra\_repro}. We extract the spectrum of the source for all obsIDs where the PSF of X-2 is not blended with those of nearby sources. Since our aim is to carry out spectral fitting, we do not use the 3 obsIDs of HRC data due to the limited spectral capabilities of this instrument. After excluding 6 obsIDs where the PSF of X-2 is blended and the 3 obsIDs of HRC data, our dataset for spectral extraction consists of 19 ACIS observations. We use the {\sc ciao} tool {\sc specextract} to extract the spectra, which produces source and background spectra and redistribution matrices and auxilliary response files (RMF and ARFs). For on-axis observations, we use circular regions of radius 1.2\arcsec\ centered on the source. A radius of 1.2\arcsec\ encircles 85\% of the energy at 6 keV for an on-axis point source. This radius increases by a factor of $\sim2$ for off-axis angles of $\sim4$\arcmin, and the PSF elongates into an elliptical shape. Therefore for off-axis observations, we use elliptical regions with 2\arcsec\ and 1\arcsec\ semi-major and semi-minor axes respectively. For observations where X-2 is faint or absent in the image, we center these regions at the same positions relative to the bright persistent source X-1 as for the obsIDs where X-2 is clearly detected. We also use smaller, 0.8\arcsec\ regions for these observations to avoid contamination from the two nearby sources to the south.

\begin{figure}
\begin{center}
\includegraphics[width=90mm]{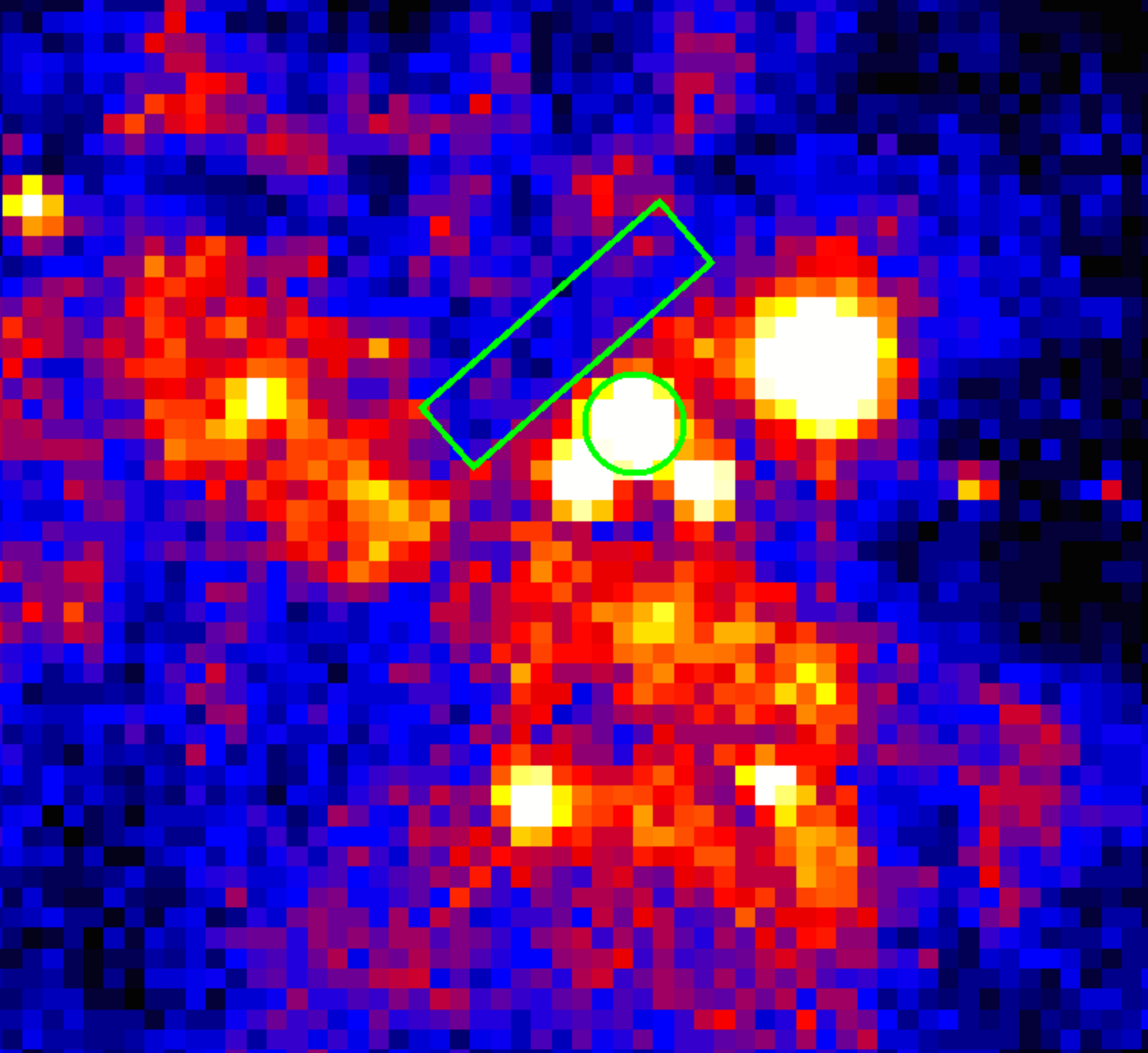}
\caption{Figure of obsID 5644 showing the source (circular) and background (rectangular) spectral extraction regions. The scaling has been modified in this figure with respect to Fig. \ref{fig_images} to emphasize the diffuse emission. The background region was chosen to contain the diffuse emission which was as local to the source as possible while also excluding nearby point sources and the PSF of X-1. North is up and east is left.}
\label{fig_regions}
\end{center}
\end{figure}

In order to carry out background extraction, we initially considered a large circular region outside the galaxy on the same chip as the galaxy, however we find that the spectra resulting from the subtraction of this background contain a prominent residual soft component. When extracting the spectra of the diffuse emission \citep[see ][]{ranalli08} from a region nearby X-2, we determine that the soft component likely results from this diffuse emission. Thus we use the spectrum of the diffuse emission for background subtraction. Figure \ref{fig_regions} shows the \chandra\ image from a 75-ks on-axis exposure of M82 (obsID 5644), where the scaling enhances the diffuse emission. This reveals the various structures present in the diffuse emission around X-2, which is particularly strong to the south east. In our choice of background region, we aim to avoid the brightest diffuse emission which is too far from the source to contribute to its background, plus to avoid the point sources to the south and the PSF of X-1 to the west. This results in a relatively small 7.6\arcsec$\times$1.9\arcsec\ rectangular extraction region to the north east of X-2, which though small contains enough signal ($\sim20$ counts ks$^{-1}$) to undertake the background extraction. We use this background region for all obsIDs.

A major effect that must be taken into account before we can perform spectral fitting is pile-up. Pile-up occurs when more than one photon lands on the same pixel between two readouts of the CCD. In this case multiple photons are counted as a single event with increased energy, or they are not counted at all (in the case of grade migration). Pile-up therefore affects both the inferred flux and spectral shape. The standard readout time for the ACIS detectors is 3.2 s, and with a count rate that can get up to $\sim$1~count~s$^{-1}$, the effect of pile-up on X-2 observations cannot be neglected. As mentioned above, the effect of pile-up can be reduced by the use of a sub-array of pixels that reduces the readout time to 0.4 s for 1/8 ACIS-S chip, or by moving the source off-axis, which spreads the photons over several pixels. Off-axis observations have the drawback that the PSF of close sources can become blended.

To estimate the level of pile-up in our data, we use the {\sc ciao} tool {\sc pileup\_map}, which outputs an image of counts per ACIS frame, which can then be used to estimate a pile-up fraction\footnote{http://cxc.harvard.edu/csc/memos/files/Davis\_pileup.pdf}. For off-axis and sub-array observations, the count rate per frame is typically less than 0.1, which corresponds to a pile-up fraction of $<5$\%. For on-axis observations with full CCDs, the counts per frame are many times higher, with typical pile-up fractions $>10$\%. We list these estimates in Table \ref{table_obsdat}. We identify 4 observations where the estimated pile-up fraction is $>10\%$, 2933, 11104, 13796 and 16580. The effect of pile-up can be accounted for in spectral fitting for cases where pile-up is not too strong, which we judge to be the case here, using a convolution model based on an algorithm presented in \cite{davis01}.

We use {\sc xspec} (v12.8.2) to carry out spectral fitting of X-2. We group the spectra with a minumum of 20 counts per bin using the {\sc heasarc} tool {\sc grppha} using the \chisq\ fit statistic and carrying out background subtraction. There are not enough counts in the spectra extracted from obsID 1302 and 15616 for \chisq\ fitting, thus we group the spectrum to have a minimum of 1 count per bin and use the Cash fit statistic. We fit the spectra in the energy range 0.5$-$8 keV with two models, a power-law ({\tt powerlaw}) model and a disk black body ({\tt diskbb}) model, both of which are subjected to photoelectric absorption local to the source ({\tt zwabs}). 

Despite subtracting off the diffuse background close to the position of X-2, we find that soft excess emission still persists over the above models. It is possible that this is residual diffuse emission not accounted for by the background subtraction, however it may also originate from optically-thin plasma that has been photo-ionized by the pulsar, evidence for which has been found around other ULXs such as Holmberg II X-1 \citep{dewangan04}, NGC 7424 ULX 2 \citep{soria06} and NGC 5408 X-1 \citep{strohmayer07}.  We account for this soft excess component in any case by adding the collisionally-ionized diffuse gas model, {\tt apec}. While many X-ray pulsars indeed present soft X-ray excesses thought to be produced by the reprocessing of the hard X-rays in the inner region of the accretion disk \citep[see e.g. ][]{hickox04}, the absorption in this system (\nh$\sim3\times10^{22}$ \cmsq) makes it unlikely that the excess we observe is intrinsic to X-2. Similarly, we find that if we subject the {\tt apec} model to the same absorption that is applied to the {\tt powerlaw} or {\tt diskbb} model, it cannot account for the soft excess. Furthermore, adding absorption to this component that is independent from the absorption of the pulsar does not produce an improvement in \chisq. We conclude that the soft excess most likely originates from foreground material in the line-of-sight that is sufficiently outside the bulk of the diffuse gas that it is not absorbed

The absorption column and $\Gamma$ (or $T_{\rm in}$) are degenerate given a limited bandpass (i.e. a hard spectrum can be fitted with a steep $\Gamma$ and high \nh\ or a flat $\Gamma$ and low \nh). For this reason we simultaneously fit for \nh\ across all epochs, which is thus driven by the spectra with the greatest counts. The parameters of the {\tt powerlaw} or {\tt diskbb} model are free for each epoch. The absorption in the spectrum is attributed to the interstellar medium in M82 rather than being local to the neutron star, and thus is not expected to change on the time-scales that we are considering. We confirm this explicitly by fitting the spectra individually with free \nh\ parameters, and find no evidence for a variable absorber. Furthermore, as we attribute the soft excess to diffuse emission, we also simultaneously fit for the temperature of the {\tt apec} model. We leave the normalizations of this model free since the small circular source extraction regions for on-axis observations and the larger elliptical extraction regions for off-axis observations contain differing amounts of the diffuse emission.

For both model cases we test the effect of adding the multiplicative {\tt pileup} model. In these cases, the frame time parameter is set dependent on the size of the sub-array used. We set the PSF fraction (not the fraction of the PSF included in the extraction region but the fraction of counts in the region which are from the point source whose pile-up is being modeled) to 95\%. The only parameter left free when using this model is $\alpha$, which is the grade morphing parameter. The parameter $\alpha$ is related to the grade migration function, $G=\alpha^{p-1}$, where p is the number of piled photons. We consider four factors when deciding whether to include this model in the spectral fit. We visually inspect the spectrum for a telltale turn-up at high energies. We use the estimated pile-up fraction from the {\sc pileup\_map} tool and consider if the observation was taken off-axis and/or with a sub-array. Generally on-axis observations show pile-up fractions greater than 10\% and we thus include the {\tt pileup} model in the fit. Lastly, if the {\tt pileup} model is included and the best-fit $\alpha$ parameter converges on a small number or zero, which is considered unphysical, we remove the {\tt pileup} model from the fit.

For the {\tt powerlaw} model, the free parameters are the photon-index, $\Gamma$, and the normalization. We allow $\Gamma$ to vary between -3 and 10. For the {\tt diskbb} model, the free parameters are $T_{\rm in}$, the temperature of the inner disk in keV, and the normalization. We restrict the disk temperature to $<10$ keV. For both models the normalization of the {\tt apec} model is a free parameter. As described above, the temperature of this component we fit for simultaneously across all epochs and found to be 0.44$^{+0.28}_{-0.25}$ keV and 0.58$^{+0.38}_{-0.17}$ keV for the {\tt powerlaw} and {\tt diskbb} models respectively. The \nh\ of the {\tt zwabs} model is also fitted simultaneously and found to be 3.4$^{+0.15}_{-0.14}\times10^{22}$ \cmsq\ and 2.8$^{+0.10}_{-0.09}\times10^{22}$ \cmsq\  respectively. The {\tt diskbb} model requires less absorption since it predicts fewer soft X-ray photons than the {\tt powerlaw} model. The redshift is fixed at 0.00067 for all model components. 

Uncertainties on the free parameters are calculated using the $\Delta$\chisq=4.61 criterion, which corresponds to 90\% confidence level for two interesting parameters. 

\section{NuSTAR data and analysis}
\label{sec_nanalysis}

\nustar\ \citep{harrison13} observed the M82 field 7 times between 2014 January 23 and 2014 March 06, as described in B14, for a total exposure of 1.91 Ms during which the pulsations were detected. A \chandra\ exposure (47.5 ks, obsID 16580) overlapped with the \nustar\ observation during a period where the pulsations were present. While B14 presented timing and photometric analysis of the pulsar, we aim here to present for the first time some of its spectral characteristics above 10 keV by isolating the pulsed component. For this we used the \nustar\ data analysis software (NuSTARDAS) version 1.2.0 and \nustar\ CALDB version 20130509 with the standard filters to obtain good time intervals, excluding the periods where the source was occulted by the Earth or was transiting through the South Atlantic Anomaly. We used the pulsar ephemeris described in B14 to extract ``pulse-on'' and ``pulse-off'' spectra. The pulse-on spectrum is defined to be the brightest 25\% of the pule profile, while the pulse-off is defined to be the faintest 25\% of the pulse profile. We then subtract the pulse-off spectrum from the pulse-on spectrum to obtain the pulsed spectrum which we model with some simple models in order to characterize the data and to facilitate comparison with other well-studied pulsars. Due to the triggered read out of the \nustar\ detectors, pile-up is not an issue at the flux levels of M82. The \nustar\ data were rebinned to a signal-to-noise of 3, providing a significant signal up to $\sim40-$50 keV. We fit the pulsed spectrum with models consisting of an absorbed power-law continuum both with and without an exponential cut-off ({\tt cutoffpl}), both of which are subjected to interstellar absorption, where we fix the \nh\ to $3\times10^{22}$ \cmsq, as determined from the \chandra\ analysis.

\section{Results}
\label{sec_results}

The results of the spectral fits using both the power-law and disk black-body model are given in Table \ref{table_specfits}. The fit statistic, \chisq, for the combined fits with the power-law model is 1439.89 with 1315 degrees of freedom (\rchisq=1.09), whereas for the disk black body model \chisq=1323.03 with 1315 degrees of freedom (\rchisq=1.01). Comparing these fit statistics suggests that the disk black body model is overall the best model. However, at the lowest luminosities ($\sim10^{38}$ \ergs), the temperature of the disk black body model is not well constrained and hits the upper bound of 10 keV imposed for spectral fitting during the error calculations. In some cases the error calculations fail. This is most likely due to the low number of counts at these luminosities.

\begin{table*}
\centering
\caption{Spectral fitting results}
\label{table_specfits}
\begin{center}
\begin{tabular}{r c c c c c c c }
\hline
ObsID & Date & \multicolumn{3}{c}{Power-law model parameters} & \multicolumn{3}{c}{Disk black body model parameters} \\
 & & $\alpha_{\rm pl}$ & $\Gamma_{\rm pl}$ & \lx$_{\rm , pl}$ & $\alpha_{\rm disk}$ & $T_{\rm in, disk}$ & \lx$_{\rm , disk}$ \\
& & & & ($\times10^{38}$ \ergs) & & (keV) & ($\times10^{38}$ \ergs) \\
(1) & (2) & (3) & (4) & (5) & (6) & (7) & (8)  \\
\hline

\input{tab_specpar2_paper.tex}

\hline
\end{tabular}
\tablecomments{Results of the X-ray spectral fitting of the 19 obsIDs where reliable spectral information on X-2 could be extracted, fitted with an absorbed power-law and absorbed disk black body models. Column (1) gives the obsID, column (2) gives the date of the observation, column (3) gives the grade morphing parameter of the pileup model when convolved with the power-law model, if this is used in the fit (`-' indicates the pileup model was not used), column (4) gives the photon index of the power-law model, column (5) gives the unabsorbed 0.5$-$10 keV luminosity of the power-law model. Column (6) gives the grade morphing parameter of the pileup model, when convolved with the disk black body model, if this is used in the fit (`-' indicates the pileup model was not used), column (7) gives the disk temperature in keV. `$u$' indicates that this parameter hit the upper limit of 10 in the spectral fit and `$l$' indicates it hit the lower limit of 0. `-' indicates where the error calculation fails. Column (8) gives the unabsorbed 0.5$-$10 keV luminosity of the disk black body model. }
\end{center}
\end{table*}

\begin{figure*}
\begin{center}
\includegraphics[width=180mm]{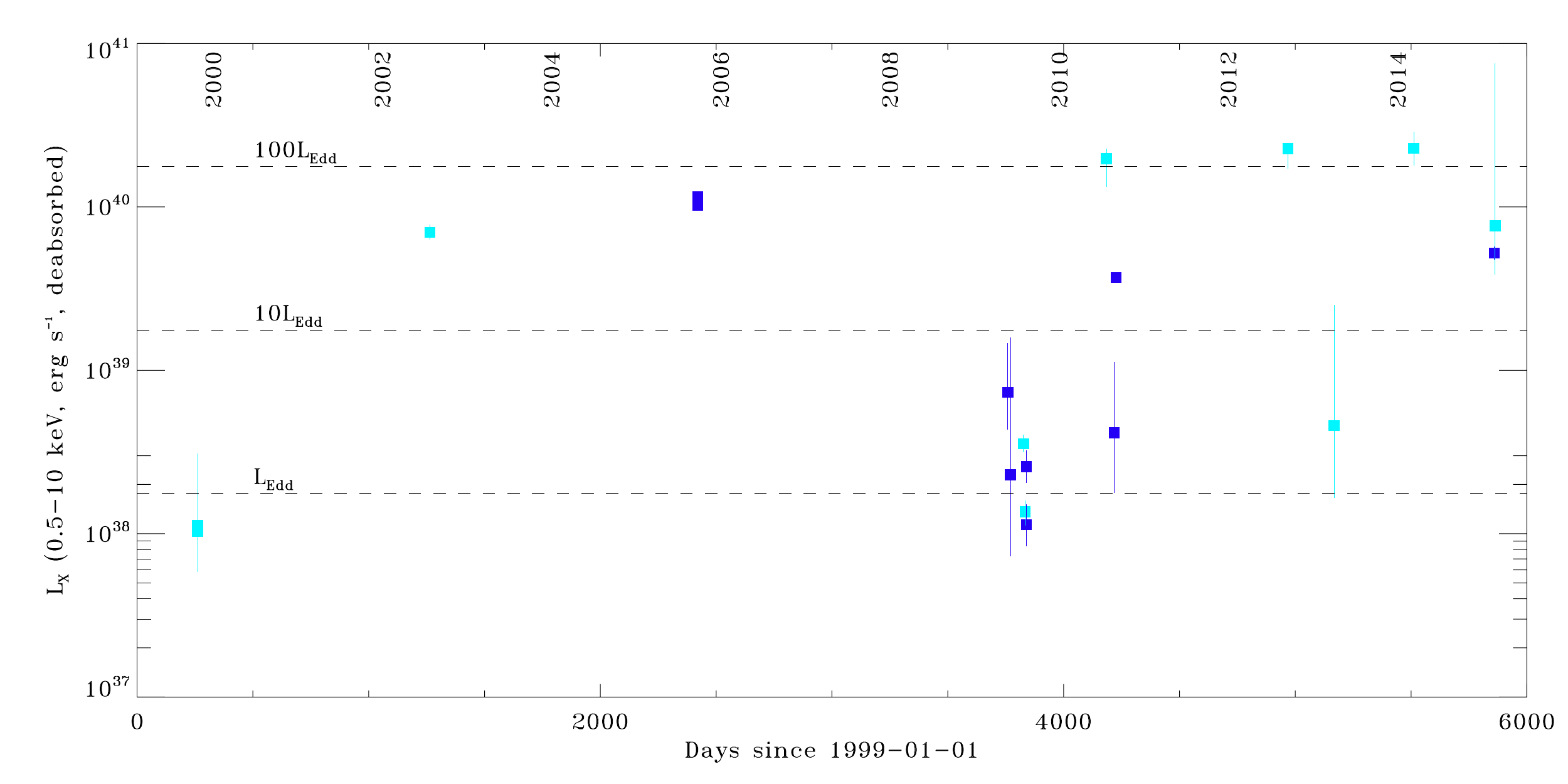}
\caption{Long-term activity of X-2 over the 15 years of \chandra\ observations showing that the source is frequently observed to be radiating at many times its Eddington limit. Vertical lines indicate the 90\% confidence range on the measured 0.5$-$10 keV luminosity, which is calculated from the power-law model assuming a distance to M82 of 3.3 Mpc. Dark blue squares show observations which were taken off-axis and/or with a sub-array of pixels to mitigate the effect of pile-up. Light blue squares are observations taken on-axis with the full array of CCDs. The horizontal lines show the Eddington limiting luminosity for a 1.4 \msol\ object, along with 10 and 100 times this.}
\label{fig_ltcrv}
\end{center}
\end{figure*}

The intrinsic luminosity estimates (corrected for absorption) between the two models differ due to both the diverging spectral shape above 8 keV (luminosity measurements are extrapolated to 10 keV for comparison with previous works) and in the soft X-ray band, plus the differing pile-up estimates caused by the difference in spectral shape of the models. The disk black body model gives a systematically lower 0.5$-$10 keV intrinsic luminosity, by typically 25\%. Nonetheless, our analysis shows that X-2 is frequently observed to be emitting well above its Eddington luminosity assuming a typical NS mass of 1.4 \msol, regardless of the model used. This is illustrated in Figure \ref{fig_ltcrv} which shows intrinsic (unabsorbed) \lx\ against time. For 9/19 (47\%) observations that we analyzed, we found that X-2 emits at a luminosity in excess of $10^{39}$ \ergs\ and thus relatively persistent, rather than transient as identified by \cite{feng07} and \cite{kong07}. 

\begin{figure}
\begin{center}
\includegraphics[width=90mm]{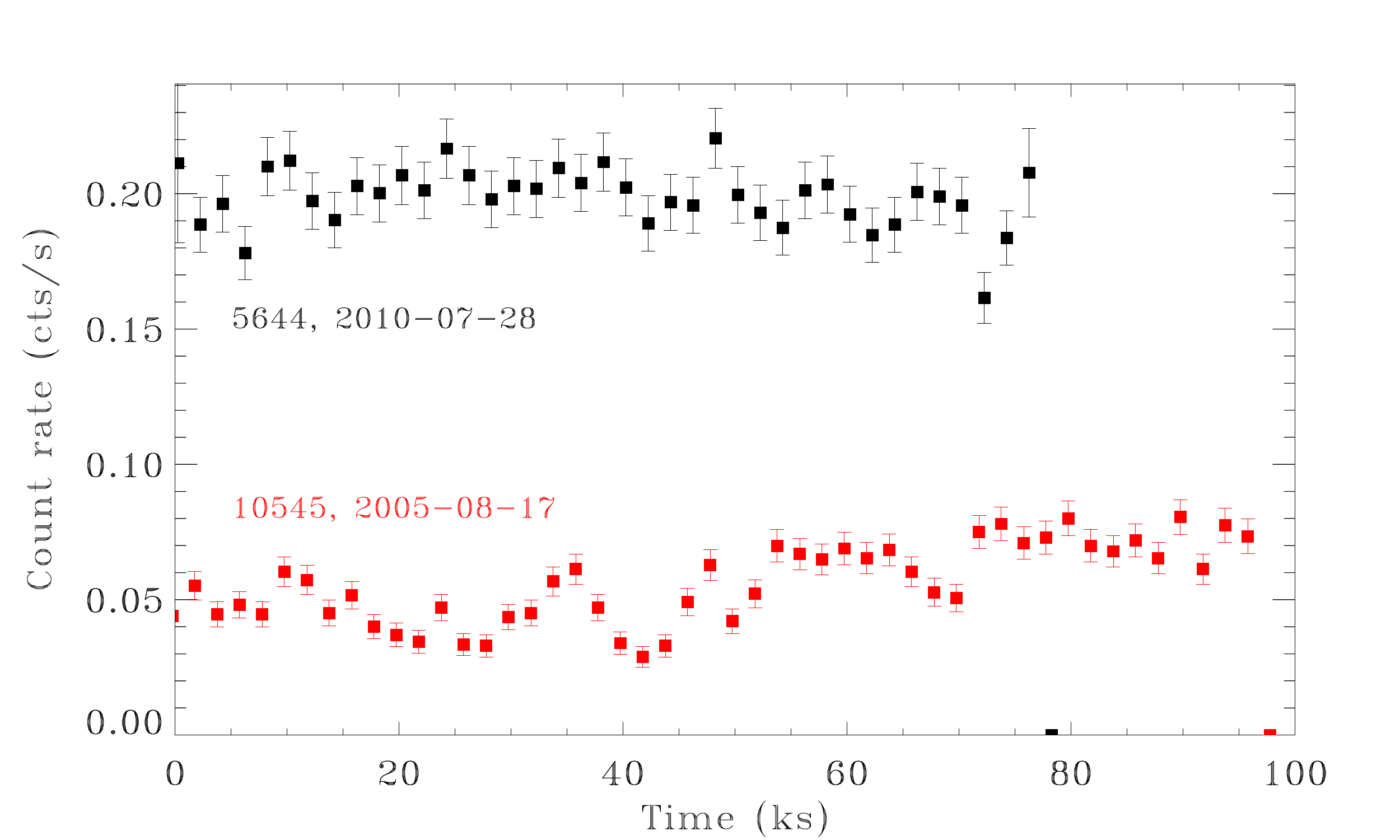}
\caption{Light-curves of X-2 for obsIDs 5644 and 10545, with 2 ks bins.}
\label{fig_ltcrv2}
\end{center}
\end{figure}

We also note the period of extreme flux variability in 2010 where the source drops almost two orders of magnitude in brightness in the space of a month. The exposure times of these two observations (11104 and 11800) are too short to reveal any significant variability during the observations, however. Of the longer observations available, obsIDs 5644 (75 ks) and 10545 (96 ks) are least affected by pile-up due to a sub array of pixels used in the former and X-2 lying at off-axis angles in the latter. The light curves of these observations are shown in Figure \ref{fig_ltcrv2}. While X-2 does not show much variability during obsID 5644, it does show variability of up to a factor of two on ks-timescales during obsID 10545.

\begin{figure}
\begin{center}
\includegraphics[width=90mm]{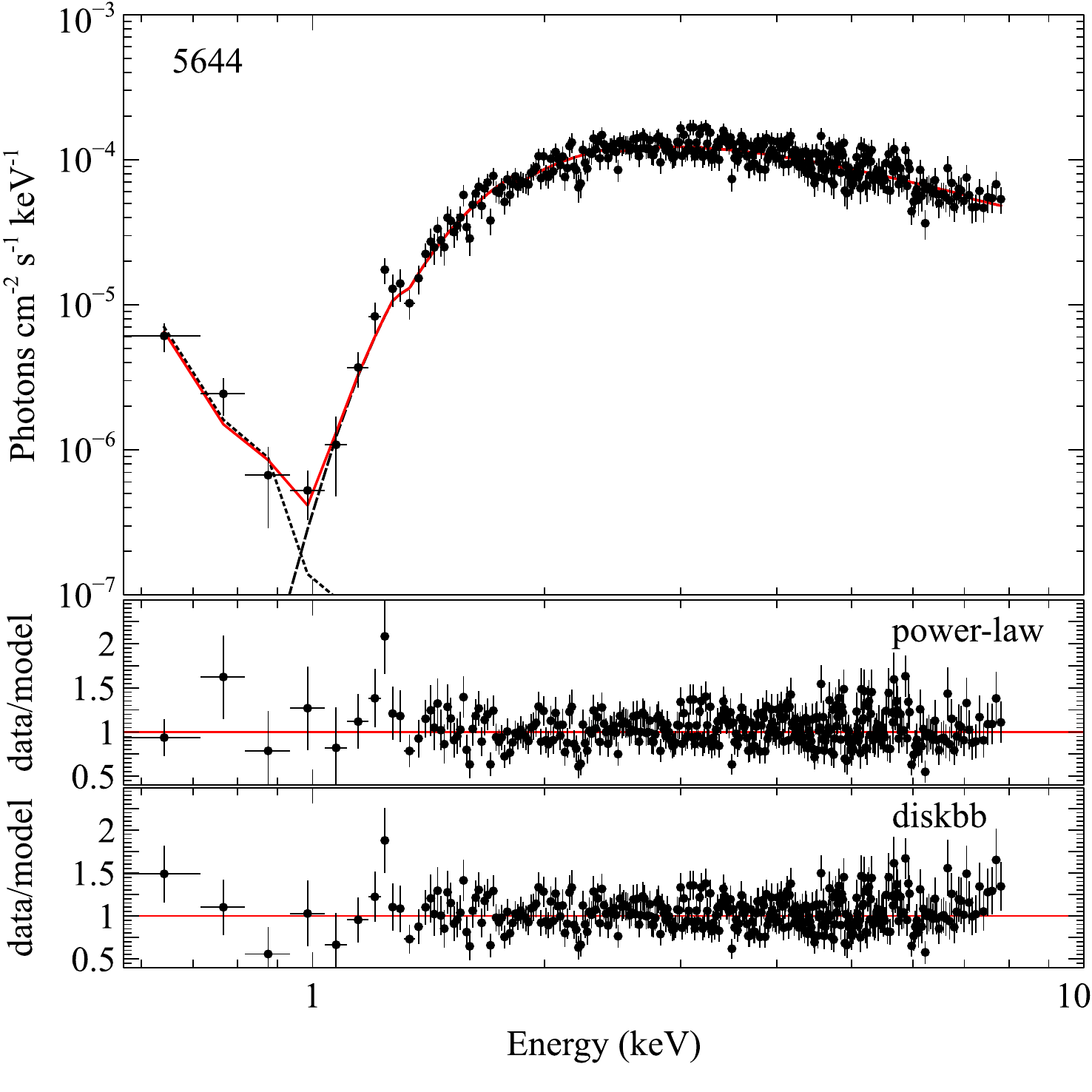}
\caption{75 ks \chandra\ spectrum of X-2 taken in 2005 (obsID 5644), taken on axis but with a sub-array of pixels to reduce pileup. The top panel shows the unfolded spectrum fitted with an absorbed power-law, plotted with a dashed line. The dotted line shows the {\tt apec} model used to model the excess diffuse background. The middle panel shows the data-to-model ratio of this fit. The bottom panel shows the data-to-model ratio of a fit with the disk black body model.}
\label{fig_spec1}
\end{center}
\end{figure}

\begin{figure}
\begin{center}
\includegraphics[width=90mm]{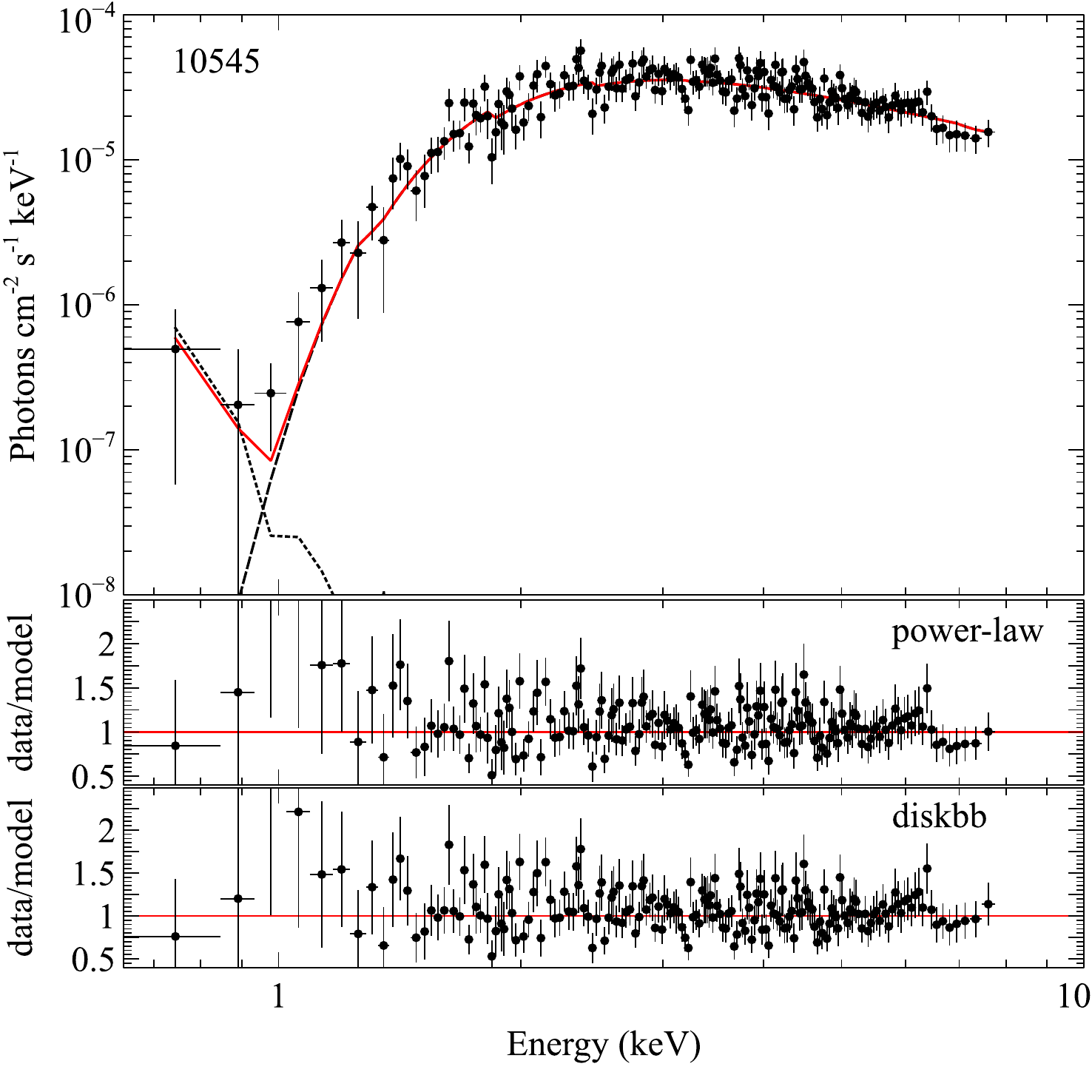}
\caption{96 ks \chandra\ spectrum of X-2 taken in 2010 (obsID 10545), taken off axis to reduce pileup. The top panel shows the unfolded spectrum fitted with an absorbed power-law, plotted with a dashed line. The dotted line shows the {\tt apec} model used to model the excess diffuse background. The middle panel shows the data-to-model ratio of this fit. The bottom panel shows the data-to-model ratio of a fit with the disk black body model.}
\label{fig_spec2}
\end{center}
\end{figure}

We also examine the spectra of X-2 for these two long observations in order to gain further insights into the source, shown in Figure \ref{fig_spec1} and Figure \ref{fig_spec2}. During obsID 5644 X-2 was observed to be near its peak luminosity of $\sim10^{40}$ \ergs, while during obsID 10545 X-2 was at a lower luminosity of $\sim4\times10^{39}$ \ergs. In Figure \ref{fig_spec1} we show the data-to-model ratios of these spectra when fitted by the power-law and disk black body models. In neither case do the ratios show any deviations indicative of a bad fit. Furthermore, the \chisq\ values are comparable between the models.

\begin{figure}
\begin{center}
\includegraphics[width=90mm]{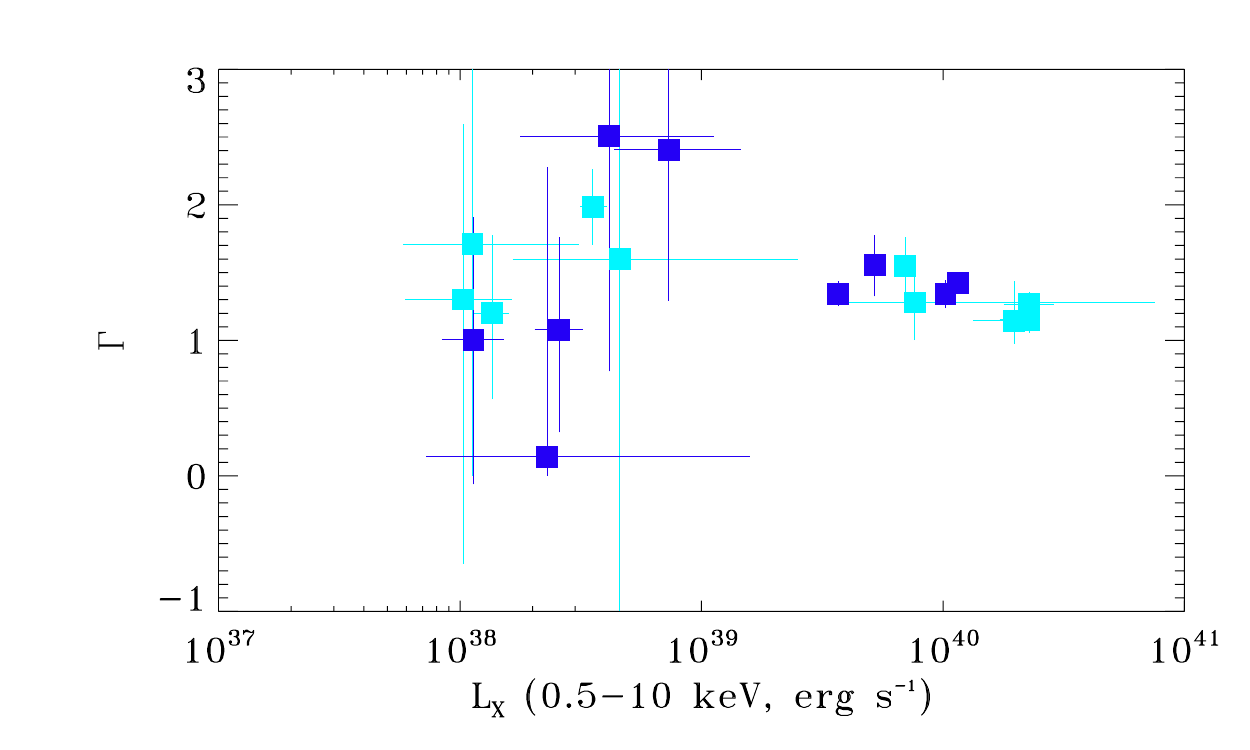}
\includegraphics[width=90mm]{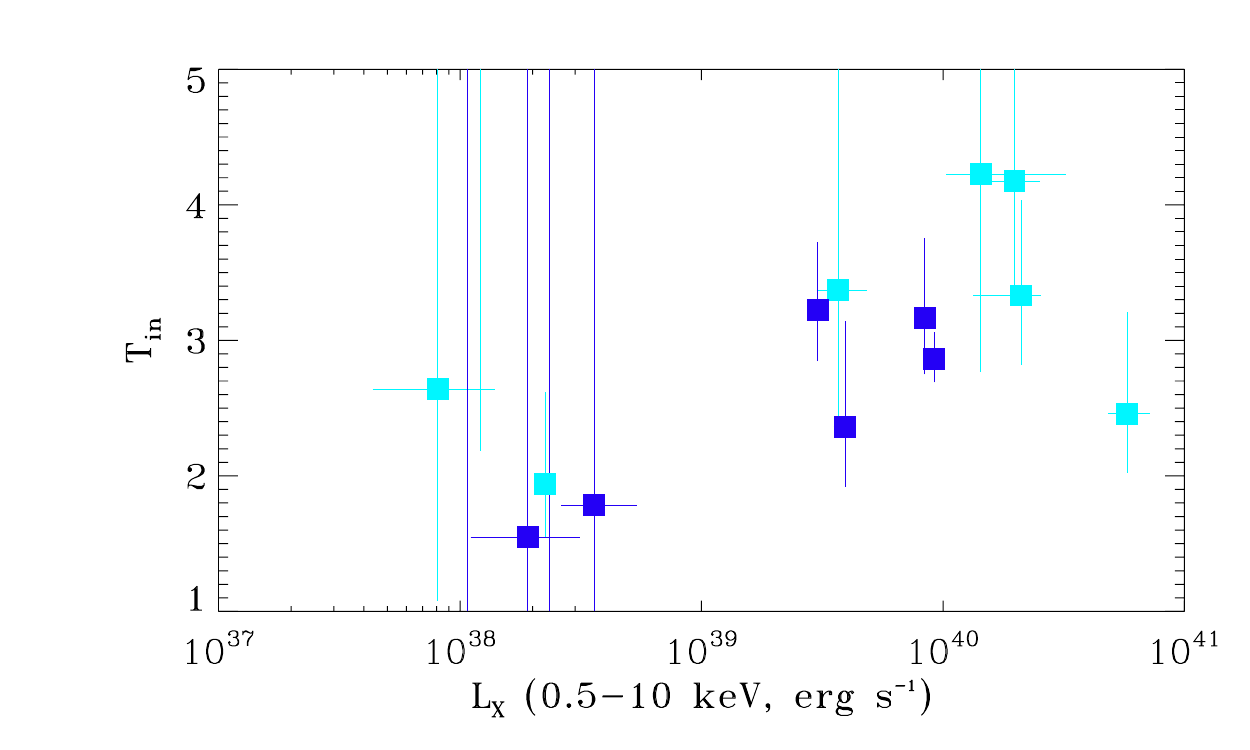}
\caption{Figures showing (top) the relationship between $\Gamma$ and \lx\ of X-2 when fitted with a power-law model and (bottom) the relationship between $T_{\rm in}$ and \lx\ when fitted with a disk black body model. Dark blue squares show observations which were taken off-axis and/or with a sub-array of pixels to mitigate the effect of pile-up. Light blue squares are observations taken on-axis with the full array of CCDs. In some cases, at the lowest luminosities, the best-fit disk temperature is off the scale of this figure and with large unconstrained values. These are indicated with vertical lines in the figure not associated with a square.}
\label{fig_specpar}
\end{center}
\end{figure}

Figure \ref{fig_specpar} shows the relationship between the spectral parameters $\Gamma$ of the power-law model and $T_{\rm in}$ of the disk black body model with respect \lx. For \lx$>10^{39}$ \ergs\ the mean $\Gamma=1.33\pm0.15$ (1$\sigma$), whereas the mean $T_{\rm in}=3.24\pm0.65$. Below 10$^{39}$ \ergs, $\Gamma$ and $T_{\rm in}$ are not well constrained and show a large spread in values.

\begin{figure}
\begin{center}
\includegraphics[width=90mm]{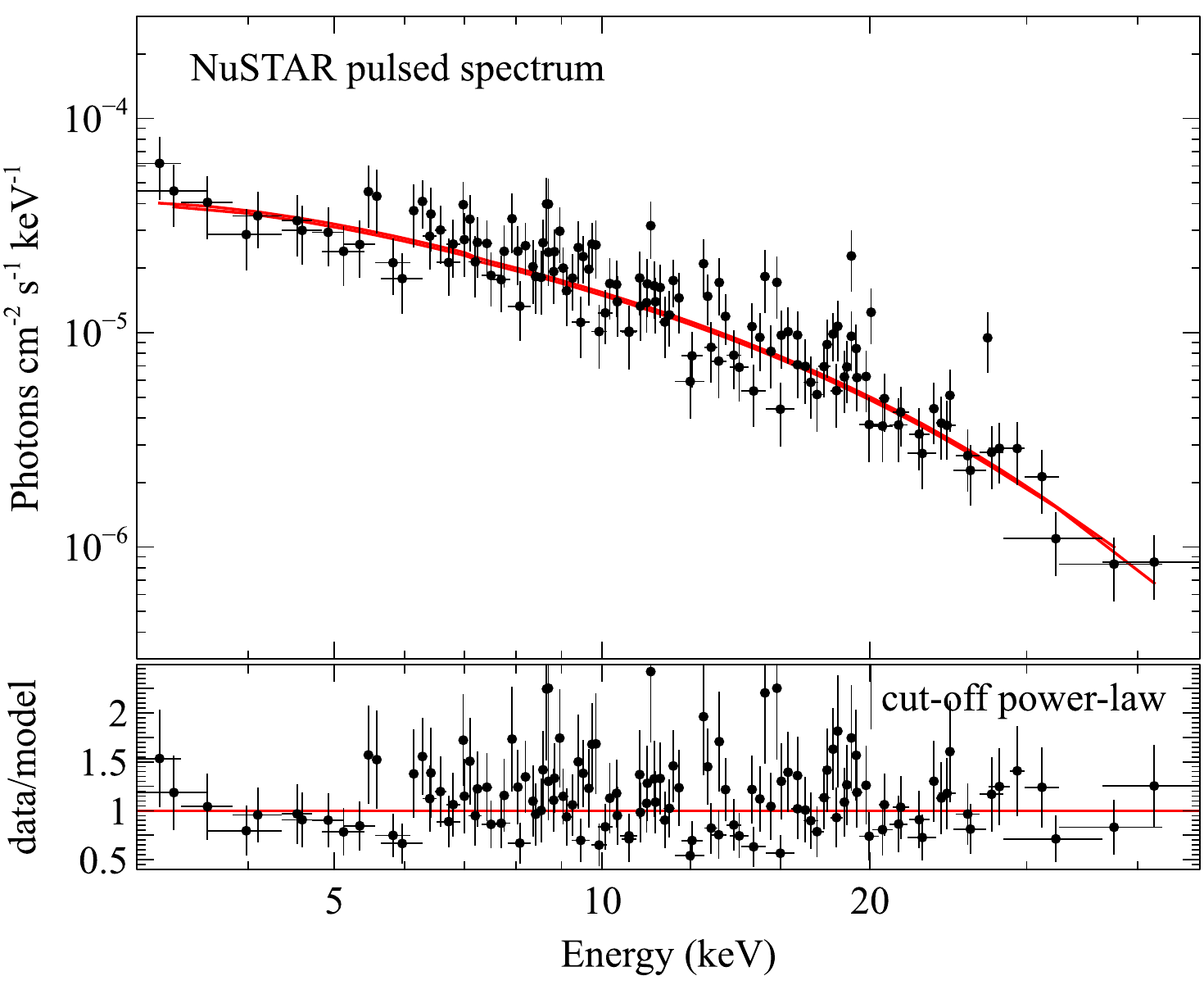}
\caption{The \nustar\ 3$-$50 keV unfolded pulsed spectrum of M82 X-2, fitted with a power-law, $\Gamma=0.6\pm0.3$, with a high energy cut-off, $E_{\rm C}=14^{+5}_{-3}$ keV. The average pulsed flux in this band is $5.7\pm0.4 \times10^{-12}$ \ergcms, corresponding to a luminosity of $7.5 \times10^{39}$ \ergs\ at 3.3 Mpc.}
\label{fig_pulsespec}
\end{center}
\end{figure}

In addition to the \chandra\ spectral analysis, we have conducted \nustar\ pulse-phased spectroscopy of the pulsar in the 3$-$50 keV range, as described in Section~\ref{sec_nanalysis}. We fit the pulsed spectrum with a power-law model and a model with an exponential cut-off ({\tt cutoffpl} in {\sc xspec}), both of which are subjected to photo-electric absorption. We find that the power-law with a cut-off is significantly preferred ($\Delta$\chisq = 20 for one additional free parameter). The fit is excellent (\chisq/DoF = 132/126) and we find the photon index of the pulsed component to be $\Gamma=0.6\pm0.3$, with a high energy cut-off, $E_{\rm C}=14^{+5}_{-3}$ keV. The average pulsed flux in this band is $5.7\pm0.4 \times10^{-12}$ \ergcms, corresponding to a luminosity of $7.5 \times10^{39}$ \ergs\ at 3.3 Mpc. We present the pulsed spectrum of M82 X-2 in Fig~\ref{fig_pulsespec}, unfolded through the instrumental response assuming the cut-off power-law model.

\section{Comparison of the M82 pulsar with other luminous X-ray pulsars and implications for theoretical models}
\label{sec_discuss}

Other examples of luminous ($>10^{38}$ \ergs) X-ray pulsars include SMC X-1, LMC X-4, GRO J1744-28, RX J0059.2-7138, XTE J0111.2-7317 and A0538-66. Although M82 X-2 can reach luminosities an order of magnitude brighter than these sources, a comparison with them is valuable. The X-ray spectra of these sources are typicaly fitted with a power-law model with $\Gamma=0.5-1.5$, subjected to absorption along the line of sight, and in some cases with a high-energy cut-off with energies ranging from 5$-$30 keV. 

\cite{paul02} fit the phase-averaged spectra of SMC X-1 and LMC X-4 in the 0.7$-$10 keV band using {\it ASCA} data. The spectrum of SMC X-1 was modeled well with a cut-off power-law, where $\Gamma=0.91\pm0.03$ and the cut-off energy is $5.5^{+1.4}_{-0.5}$ keV. The spectrum of LMC X-4 did not require a cut-off, and could be reproduced by a plain power-law with $\Gamma=0.69\pm0.04$. The other examples of super-Eddington pulsars, GRO J1744-28, RX J0059.2-7138, XTE J0111.2-7317 and A0538-66 are also well described by cut-off power-laws or power-law models, both with similar parameters \citep[see ][repectively]{nishiuchi99, younes15, sidoli15, yokogawa00, skinner82}.

The phase-averaged 0.5$-$8 keV X-ray spectral properties of M82 X-2 are very similar to these other sources, with $\Gamma=1.33\pm0.15$ for a power-law model with no cut-off. The long \chandra\ observation obsID 5644 where the source is caught in an ultra-luminous state, but where pile-up is negligible, offers us the opportunity to test for the presence of a cut-off and if any constraints can be placed on it. Fitting this spectrum with this model yields $\Gamma=0.70^{+0.68}_{-0.65}$ and $E_{\rm C}=6.19^{+50.9}_{-2.9}$ keV, where \chisq=389.87 for 338 degrees of freedom. The fit with the power-law without a cut-off gives \chisq=395.66 for 339 degrees of freedom, thus the addition of the cut-off yields an improvement in \chisq\ of only 6 for the addition of one parameter and the cut-off energy is not well constrained in the \chandra\ data alone.

\nustar, however, can measure the cut-off due to its sensitivity above 10 keV. From analysis of the pulsed component in the 3$-$50 keV range we find that the pulsed spectrum is best fitted by a power-law with a high-energy cut-off, where $\Gamma=0.6\pm0.3$ and $E_{\rm C}=14^{+5}_{-3}$ keV. These values are very similar to the phase-averaged spectrum described above, albeit with much better constraints owing to the high-energy sensitivity of \nustar.  The similarity with the phase-averaged spectrum from \chandra\ indicates that the pulsed spectrum dominates the emission of the pulsar. It is also interesting to note that the cut-off energy of the pulsed component is higher than that observed in other ULXs where the nature of the accretors remain unknown, whose spectra typically cut off at 6-8 keV \citep[e.g.][]{bachetti13, walton13, rana15, mukherjee15}

\cite{dallosso15} note the empirical relationship between the energy of the cyclotron resonance features, $E_{\rm cyc}$, of four X-ray pulsars and the cut-off energy in their spectra, $E_{\rm C}$ \citep[see ][]{makishima90}, where $E_{\rm cyc}=(1.4-1.8)\times E_{\rm C}$. Many of the theoretical modeling papers that have aimed to explain the nature of the ultra-luminous pulsar have presented different scenarios for the strength of its magnetic field. Since $E_{\rm cyc}$ is directly proportional to the strength of the magnetic field, then $E_{\rm C}$ could potentially yield information about the magnetic field strength. Following this, the 14 keV cut-off that we measure in the pulsed spectrum implies a $B\sim10^{12}$ G magnetic field. However, we note that there are exceptions to the above relation, for example KS~1947+319, where $E_{\rm cyc}$=12 keV and $E_{\rm C}$=22 keV \citep{fuerst14}. 

\cite{dallosso15} also discuss the variability exhibited by X-2, which can be explained by relatively small changes in the mass accretion rate in the presence of a strong magnetic field, whereby the source at low luminosities enters the propeller regime. We note, however, that the minimum luminosity at which accretion is possible in the presence of a $B=10^{13}$ G magnetic field, a strength which they favor, with a 1.37 s period is $\sim2\times10^{39}$ \ergs\ \citep[see ][]{stella86}. Considering that we observe X-2 at much lower luminosities, argues against such a strong magnetic field given the discussion above.

Concerning the duty cycle of X-2, we have found that the source is more persistent than previously reported by \cite{feng07} and \cite{kong07}, due to the longer baseline of our investigation and larger data set. For 9/19 (47\%) observations that we analyzed, we found that X-2 emits at a luminosity in excess of $10^{39}$ \ergs. Luminosities of $\sim10^{39}-10^{40}$ \ergs\ imply that the neutron star is growing at a rate of $\sim2\times10^{-8}-10^{-7}$ \msol\ yr$^{-1}$, assuming isotropic emission and a mass-to-energy conversion efficiency of unity, meaning it will collapse into a black hole within $\sim10-100$ million years. These results could have important implications for the formation and growth of supermassive black holes, theoretical modeling of which often employs an early super-critical accretion phase to explain the masses of the super-massive black holes found in quasars at $z\sim7$ \citep[e.g.][]{volonteri06, mortlock11}. Low-mass X-ray binaries have also been proposed to be a potential source of ionizing radiation for heating the intergalactic medium during the epoch of reionization \citep{fragos13}, which may indeed have a contribution from a non-negligible population of ultra-luminous pulsars.

\section{Summary and Conclusions}

In this paper we have conducted a temporal and spectral analysis of the 0.5-8 keV X-ray emission from the ultra-luminous X-ray pulsar M82 X-2 from 15 years of \chandra\ observations and pulse-phased spectroscopy in the 3$-$50 keV band from \nustar\ data. Our main findings are as follows:

\begin{itemize}

\item When fitted with a power-law model, the average photon index for epochs where \lx$>10^{39}$ \ergs\ in the 0.5$-$8 keV band is $\Gamma=1.33\pm0.15$. For the disk black body model, the average temperature is $T_{\rm in}=3.24\pm0.65$ keV. This spectral shape is consistent with other luminous X-ray pulsars. We also investigated the inclusion of a soft excess component and spectral break finding that the spectra are also consistent with these features common to luminous X-ray pulsars.

\item The pulsed emission of X-2 in the 3$-$50 keV band from \nustar\ data is best fitted by a power-law with a high-energy cut-off, where $\Gamma=0.6\pm0.3$ and $E_{\rm C}=14^{+5}_{-3}$ keV with a luminosity of $7.5 \times10^{39}$ \ergs.

\item Our results show that X-2 has been remarkably active over the 15-year period considered. We find that for 9/19 (47\%) observations that we analyzed, the pulsar appears to be emitting at a luminosity in excess of $10^{39}$ \ergs, which is greater than 10 times its Eddington limit. Luminosities of $\sim10^{39}-10^{40}$ \ergs\ imply that the neutron star is growing at a rate of $\sim2\times10^{-8}-10^{-7}$ \msol\ yr$^{-1}$ and is expected to collapse into a black hole within $\sim10-100$ million years.

\end{itemize}

\section{Acknowledgements}

This work made significant use of archival observations made by the \chandra\ X-ray observatory, for which we thank the builders and operators, as well as the software package {\sc ciao}. The data were obtained from the High Energy Astrophysics Science Archive Research Center (HEASARC), which is a service of the Astrophysics Science Division at NASA/GSFC and the High Energy Astrophysics Division of the Smithsonian Astrophysical Observatory. This work also made use of data from the NuSTAR mission, a project led by the California Institute of Technology, managed by the Jet Propulsion Laboratory, and funded by NASA. We thank the NuSTAR Operations, Software and Calibration teams for support with the execution and analysis of these observations. This research has made use of the NuSTAR Data Analysis Software (NuSTARDAS) jointly developed by the ASI Science Data Center (ASDC, Italy) and the California Institute of Technology (USA). AZ acknowledges funding from the European Research Council under the European Union's Seventh Framework Programme (FP/2007-2013)/ERC Grant Agreement n. 617001.

{\it Facilities:} \facility{\chandra\ (ACIS), \nustar}

\bibliography{bibdesk.bib}

\end{document}

%% file: tab_obsdat.tex
361		& 1999-09-20	& ACIS-I	& 33.7	& $<$1\%	&	on axis \\
1302		& 1999-09-20	& ACIS-I 	& 15.7	& $<$1\%	&	on axis \\
1411*		& 1999-10-28	& HRC	& 54.0      &  -	&	\\
378*		& 1999-12-30	& ACIS-I	& 4.2		& -	&	off axis, sub-array, X-2 PSF blended \\
379*		& 2000-03-11	& ACIS-I	& 9.1		& -	&	off axis, sub-array, X-2 PSF blended \\
380*		& 2000-05-07	& ACIS-I	& 5.1		& -	&	off axis, X-2 PSF blended \\
2933		& 2002-06-18	& ACIS-S	& 18.3	& $>$10\%	&	on axis \\
6097*		& 2005-02-04	& ACIS-S & 58.2	& -	&	off axis, sub-array, X-2 PSF blended \\
5644		& 2005-08-17	& ACIS-S	& 75.1	& $<$5\%	&	on axis, sub-array \\
6361		& 2005-08-18	& ACIS-S	& 19.2	& $<$5\%	&	on axis, sub-array \\
8189*		& 2007-01-09	& HRC  	& 61.6	&  - \\
8505*		& 2007-01-12	& HRC	& 83.6     &  - \\
8190*		& 2007-06-02 	& ACIS-S	& 5.8		& -	&	 off axis, X-2 PSF blended \\
10027*	& 2008-10-04	& ACIS-S	& 20.2	& -	&	off axis, X-2 PSF blended \\
10025	& 2009-04-17	& ACIS-S	& 19.2	& $<$1\%	&	off axis, sub-array \\
10026	& 2009-04-29	& ACIS-S & 18.7	& $<$1\%	&	off axis \\
10542	& 2009-06-24	& ACIS-S	& 120      & $<$1\%	&	on axis \\
10543	& 2009-07-01	& ACIS-S	& 120      & $<$1\%	&	on axis \\
10925	& 2009-07-07	& ACIS-S	& 45.1	& $<$1\%	&	off axis \\
10544	& 2009-07-07	& ACIS-S	& 74.5	& $<$1\%	&	off axis \\
11104	& 2010-06-17	& ACIS-S	& 10.1	& $>$10\%	&	on axis \\
11800	& 2010-07-20	& ACIS-S	& 17.1	& $<$1\%	&	off axis \\
10545	& 2010-07-28	& ACIS-S	& 96.3	& $\sim$5\%	&	off axis \\
13796	& 2012-08-09	& ACIS-S	& 20.1	& $>$10\%	&	on axis \\
15616	& 2013-02-24	& ACIS-S	& 2.1		& $<$5\%	&	on axis, short observation (2 ks) \\
16580	& 2014-02-03	& ACIS-S	& 47.5	& $>$10\%	&	on axis \\
17578	& 2015-01-16	& ACIS-S	& 10.1	& $\sim$1\%	&	off axis, sub-array \\
16023	& 2015-01-20	& ACIS-S	& 10.1	& $<$10\%	&	on axis \\

%% file: tab_specpar2_paper.tex
     361 & 1999-09-20 &  - &  1.30$^{+ 1.30}_{- 1.95}$ &   1.0$^{+  0.6}_{-  0.4}$ &  - &  2.64$^{+u}_{- 1.56}$ &   0.8$^{+  0.6}_{-  0.4}$ \\
    1302 & 1999-09-20 &  -&  1.71$^{+ 1.53}_{- 1.71}$ &   1.1$^{+  2.0}_{-  0.5}$ &  - &  - & - \\
    2933 & 2002-06-18 &  0.69$^{+3.1}_{- 0.39}$ &  1.55$^{+ 0.21}_{- 0.20}$ &    69$^{+    7}_{-    6}$ &  0.22$^{+ 0.14}_{- 0.11}$ &  2.46$^{+ 0.75}_{- 0.44}$ &   579$^{+  142}_{-   97}$ \\
    5644 & 2005-08-17 &  - &  1.42$^{+ 0.05}_{- 0.05}$ &   115$^{+    2}_{-    2}$ &  - &  2.86$^{+ 0.20}_{- 0.17}$ &    92$^{+    2}_{-    2}$ \\
    6361 & 2005-08-18 &  - &  1.34$^{+ 0.11}_{- 0.11}$ &   102$^{+    4}_{-    4}$ &  - &  3.16$^{+ 0.59}_{- 0.41}$ &    83$^{+    5}_{-    4}$ \\
   10025 & 2009-04-17 &  - &  2.41$^{+ 0.90}_{- 1.11}$ &   7.3$^{+  7.3}_{-  3.0}$ &  - &  1.78$^{+ u}_{- 0.84}$ &   3.6$^{+  1.8}_{-  1.0}$ \\
   10026 & 2009-04-29 &  - &  0.14$^{+ 2.14}_{- 0.14}$ &   2.3$^{+ 13.6}_{-  1.6}$ &  - & - &   - \\
   10542 & 2009-06-24 &  - &  1.99$^{+ 0.28}_{- 0.28}$ &   3.5$^{+  0.5}_{-  0.4}$ &  - &  1.94$^{+ 0.67}_{- 0.40}$ &   2.3$^{+  0.1}_{-  0.2}$ \\
   10543 & 2009-07-01 &  - &  1.20$^{+ 0.57}_{- 0.64}$ &   1.4$^{+  0.2}_{-  0.2}$ &  - &  5.21$^{+ u}_{- 3.03}$ &   1.2$^{+  0.3}_{-  0.3}$ \\
   10925 & 2009-07-07 &  - &  1.08$^{+ 0.68}_{- 0.76}$ &   2.6$^{+  0.7}_{-  0.5}$ &  - &  7.42$^{+ u}_{- l}$ &   2.3$^{+  0.5}_{-  0.7}$ \\
   10544 & 2009-07-07 &  - &  1.00$^{+ 0.91}_{- 1.07}$ &   1.1$^{+  0.4}_{-  0.3}$ &  - &  9.01$^{+ u}_{- l}$ &   1.1$^{+  0.3}_{-  0.4}$ \\
   11104 & 2010-06-17 &  0.20$^{+ 0.29}_{- 0.17}$ &  1.14$^{+ 0.29}_{- 0.17}$ &   197$^{+   29}_{-   64}$ &  0.27$^{+ 0.33}_{- 0.24}$ &  4.23$^{+ u}_{- 1.46}$ &   143$^{+  180}_{-   40}$ \\
   11800 & 2010-07-20 &  - &  2.51$^{+ 1.25}_{- 1.74}$ &   4.1$^{+  7.1}_{-  2.4}$ &  - &  1.55$^{+ u}_{- 0.88}$ &   1.9$^{+  1.2}_{-  0.8}$ \\
   10545 & 2010-07-28 &  - &  1.34$^{+ 0.09}_{- 0.09}$ &    36$^{+    1}_{-    1}$ &  - &  3.22$^{+ 0.51}_{- 0.38}$ &    30$^{+    1}_{-    1}$ \\
   13796 & 2012-08-09 &  0.24$^{+ 0.14}_{- 0.09}$ &  1.15$^{+ 0.19}_{- 0.10}$ &   227$^{+    0}_{-   55}$ &  0.25$^{+ 0.12}_{- 0.09}$ &  4.17$^{+ 1.38}_{- 0.83}$ &   197$^{+   55}_{-   50}$ \\
   15616 & 2013-02-24 &  - &  1.60$^{+ 2.48}_{- 3.18}$ &   4.6$^{+ 20.5}_{-  2.9}$ &  - &  - & - \\
   16580 & 2014-02-03 &  0.27$^{+ 0.12}_{- 0.06}$ &  1.27$^{+ 0.09}_{- 0.09}$ &   227$^{+   61}_{-   49}$ &  0.27$^{+ 0.20}_{- 0.05}$ &  3.33$^{+ 0.70}_{- 0.51}$ &   210$^{+   43}_{-   77}$ \\
   17578 & 2015-01-16 &  - &  1.55$^{+ 0.22}_{- 0.23}$ &    52$^{+    4}_{-    4}$ &  - &  2.36$^{+ 0.78}_{- 0.45}$ &    39$^{+    4}_{-    4}$ \\
   16023 & 2015-01-20 &  - &  1.28$^{+ 0.27}_{- 0.28}$ &    76$^{+  678}_{-   38}$ &  - &  3.37$^{+ 3.55}_{- 1.04}$ &    36$^{+   11}_{-    6}$ \\